\documentclass[aps,prl,reprint,balancelastpage,nofootinbib,preprintnumbers,superscriptaddress]{revtex4-1}

\usepackage{slashed}
\usepackage{bm}
\usepackage{latexsym,amssymb,amsmath,float,url,mathrsfs}
\usepackage{latexsym}
\usepackage{graphicx}
\usepackage{epstopdf}
\usepackage{amsmath}
\usepackage{subfigure}
\usepackage{natbib}
\usepackage{hyperref}
\usepackage{pifont}
\usepackage{blindtext}
\usepackage{multirow}
\usepackage{tabularx}
\usepackage[table]{xcolor}
\usepackage{flushend}

\usepackage{graphics, appendix,afterpage,makecell}

\newcolumntype{P}[1]{>{\centering\arraybackslash}p{#1}}

\newcommand\Tstrut{\rule{0pt}{2.6ex}}         
\newcommand\Bstrut{\rule[-0.9ex]{0pt}{0pt}}   

\hypersetup{
     colorlinks   = true,
     citecolor    = violet,
     urlcolor     = violet,
     linkcolor    = violet
}

\begin{document}

\title{Spurious Point Source Signals in the Galactic Center Excess}
\preprint{MIT-CTP/5170}

\author{Rebecca K. Leane}
\thanks{{\scriptsize Email}: \href{mailto:rleane@mit.edu}{rleane@mit.edu}; {\scriptsize ORCID}: \href{http://orcid.org/0000-0002-1287-8780}{0000-0002-1287-8780}}
\affiliation{Center for Theoretical Physics, Massachusetts Institute of Technology, Cambridge, MA 02139, USA}

\author{Tracy R. Slatyer}
\thanks{{\scriptsize Email}: \href{mailto:tslatyer@mit.edu}{tslatyer@mit.edu}; {\scriptsize ORCID}:
\href{http://orcid.org/0000-0001-9699-9047}{0000-0001-9699-9047}}
\affiliation{Center for Theoretical Physics, Massachusetts Institute of Technology, Cambridge, MA 02139, USA}

\date{\today}

\begin{abstract} 
We re-examine evidence that the Galactic Center Excess (GCE) originates primarily from point sources (PSs). We show that in our region of interest, non-Poissonian template fitting (NPTF) evidence for GCE PSs is an artifact of unmodeled north-south asymmetry of the GCE. This asymmetry is strongly favored by the fit (although it is unclear if this is physical), and when it is allowed, the preference for PSs becomes insignificant. We reproduce this behavior in simulations, including detailed properties of the spurious PS population. We conclude that NTPF evidence for GCE PSs is highly susceptible to certain systematic errors, and should not at present be taken to robustly disfavor a dominantly smooth GCE.
\end{abstract}

\maketitle

\noindent

Data from the \textit{Fermi} Gamma-Ray Space Telescope have revealed an intriguing excess of GeV-scale gamma rays from the region around the Galactic Center~\cite{Goodenough:2009gk, Hooper:2010mq, Hooper:2011ti, TheFermi-LAT:2015kwa}. The origin of this Galactic Center Excess (GCE) has been an active controversy for some years, with much interest in the possibility that it might be the first detected signal of annihilating dark matter (DM). In 2015, two papers made data-driven arguments that the GCE was likely to represent a previously-undetected population of point sources (PSs) in the inner Galaxy, most likely pulsars~\cite{Bartels:2015aea, Lee:2015fea}; subsequent analyses have argued for a stellar origin for the GCE based on the signal morphology~\cite{Macias:2016nev, Bartels:2017vsx, Macias:2019omb}. All these analyses are subject to systematic uncertainties from modeling of non-GCE gamma rays, but they had led to a general view that the GCE is unlikely to be explained by a diffuse source of gamma-ray emission (such as DM annihilation). However, this consensus has recently been challenged \cite{Leane:2019xiy,Zhong:2019ycb}, reigniting the debate on the origin of the GCE.

In this work we reassess the apparent evidence for a PS-dominated GCE from Non-Poissonian Template Fitting (NPTF) methods, as introduced in Ref.~\cite{Lee:2015fea}. This approach models the gamma-ray sky as a linear combination of spatial templates describing various contributions to the  gamma-ray flux. Templates provide the expected flux in each spatial pixel as a function of model parameters; the probability of obtaining a specified number of photons in the pixel can be described by either a Poisson distribution or by non-Poissonian statistics. The Poisson distribution is appropriate when the photon events are independent, and is relevant for diffuse emission and point sources where the locations of the sources are known. If the individual positions of point sources are unknown, then a non-Poissonian distribution is required, to account for the number of sources in the pixel being variable~\cite{Malyshev:2011zi,Lee:2014mza}. We will loosely refer to these two cases as ``smooth'' and ``point-like'' / ``PS'' templates respectively, although templates with Poissonian statistics can still have sharp variations in the expected flux from pixel to pixel. Given a set of templates with associated photon statistics (Poissonian or non-Poissonian), the likelihood for the observed data can be computed, and then used to generate posterior probability distributions for the model parameters. Ref.~\cite{Lee:2015fea} found a strong statistical preference for a GCE PS template with flux sufficient to explain the entire GCE, and interpreted this as evidence for a new GCE-correlated PS population.

In this \textit{Letter} we will explicitly demonstrate that the NPTF preference for PSs can change dramatically as a result of a simple perturbation to the signal model. Working in a $10^\circ$ radius region of interest (ROI), we show that when the northern and southern halves of the GCE are allowed to float independently, their coefficients are asymmetric at high significance. We find that including this asymmetry in the model removes the apparent evidence for GCE PSs, and that simulating north-south asymmetry but not allowing it in the analysis creates a spurious preference for GCE PSs. We discuss the generality of our results in a companion paper~\cite{PhysRevD.102.063019}.
\\

\noindent
\textbf{\textit{Methodology and Data Selection.}}

To employ the NPTF method, we use the NPTF package \texttt{NPTFit} \cite{Mishra-Sharma:2016gis}, interfaced with the Bayesian interference tool \texttt{MultiNest}~\cite{Feroz:2008xx}. The total number of live points for all \texttt{MultiNest} runs is \mbox{\texttt{nlive = 500}} unless specified otherwise. Mock data for PS populations are generated using \texttt{NPTFit-Sim}~\cite{NPTFSim}. We use the \texttt{Pass 8} \textit{Fermi} data, in the energy range $2-20$ GeV and collected over 573 weeks, from August 4th 2008 to June 19th 2019. We employ only events from the \textsc{UltracleanVeto} (1024) class, which has the most stringent cosmic-ray rejection cuts. This is further restricted to the top three quartiles of events graded by angular reconstruction (PSF1$-$PSF3), with quality cuts \texttt{DATA\_QUAL==1 \&\& LAT\_CONFIG==1}.  The maximum zenith angle is $90^\circ$. Compared to our previous work~\cite{Leane:2019xiy}, we have increased the photon dataset both by updating in time ($\sim 3$ additional years) and by inclusion of the top three quartiles, rather than only the top quartile.

Our modeling includes Poissonian templates for the Galactic diffuse emission, isotropic emission (``Iso''), emission in the \textit{Fermi} Bubbles (``Bub''), and the GCE (``GCE Smooth''), and non-Poissonian templates for PSs tracing the Galactic disk (``Disk PS''), isotropic emission (``Iso PS''), and the GCE (``GCE PS''). These are the same templates defined in Ref.~\cite{Leane:2019xiy} (up to a factor of the difference in exposure between the datasets), although the templates we label as ``GCE Smooth'' and ``GCE PS'' are labeled respectively as ``NFW DM'' and ``NFW PS'' in that work. ``NFW'' stands for Navarro-Frenk-White~\cite{Navarro:1995iw}, a commonly-employed prescription for the DM density profile; we use a generalized NFW profile with inner slope 1.25, which has previously been found to be a good description of the GCE. For GCE PSs, the assumed PS distribution is technically NFW$^2$, to match the morphology of an annihilating DM signal.

For the diffuse model, we use the \textit{Fermi} \texttt{p6v11} diffuse model as a baseline (used to claim evidence for PSs in Ref.~\cite{Lee:2015fea}), and check results also with the GALPROP-based~\cite{Strong:1998pw} Galactic diffuse emission models denoted \texttt{Model A} and \texttt{Model F} in Ref.~\cite{Calore:2014xka}. We do not employ the more recent \textit{Fermi} Pass 7 and Pass 8 diffuse models because choices were made in their construction that render them unsuited for studies of extended diffuse emission (see \cite{Leane:2019xiy} for further discussion).  When testing the spatial morphology of various components, templates may be broken into sub-regions, as described below. For all non-Poissonian templates, we assume a broken power-law form for the source count function (SCF) describing the number of sources as a function of flux.  Our priors are detailed in the Supplemental Material; unless specified otherwise, all components are restricted to have non-negative fluxes.

 When testing for the presence of GCE PSs, we compute the ratio of Bayesian evidences for models with and without the GCE PS template; this Bayes factor describes the strength of evidence in favor of GCE-distributed PSs.

The ROI  for our study is within a 10$^\circ$ radius circle of the Galactic Center, excluding the $|b|< 2^\circ$ band along the Galactic plane. 
We choose this ROI as previous studies have found the GCE extends out to at least 10$^\circ$ from the Galactic Center~\cite{Daylan:2014rsa}, but at the same time there is reason to think that systematic effects from the mis-modeling of the Galactic diffuse emission are likely to be less severe in smaller ROIs~\cite{bennick,Daylan:2014rsa,Chang:2018bpt}. We examine additional ROIs in our companion paper~\cite{PhysRevD.102.063019}.
\\

\noindent
\textbf{\textit{Breaking up the GCE Removes Evidence for Point Sources.}}

We begin by subdividing both GCE Smooth and GCE PS templates into northern and southern components, and allow all parameters of these four templates to float independently in a fit to the real data.

Figure~\ref{fig:a_flux} shows the flux attributed to each template. The flux attributed to the GCE PS templates is consistent with zero, and instead the flux posteriors for the GCE Smooth components are peaked at positive values.

Table~\ref{tab:bayesNFWreg} shows the preference for PSs in the analyses with and without subdivision of the GCE templates. We find that floating both GCE PS and GCE Smooth templates with a single coefficient over the whole ROI, as has been done in past NPTF papers, yields strong apparent evidence for PSs, with a Bayes factor of $\sim10^{15}$ when using the \texttt{p6v11} Galactic diffuse emission model. Yet, when the GCE templates are broken up into north/south components, the preference for PSs is essentially entirely lost, with Bayes factors $< 10$; thus even a large PS preference may be solely driven by unmodeled asymmetry. Testing other diffuse models, we find that any preference for PSs is removed once the asymmetry is allowed, although a large preference for PSs is in any case not ubiquitous (e.g. \texttt{Model F} finds no preference for PSs). For additional comparisons, see the Supplemental Material.

This serves as an explicit existence proof that the NPTF method can be highly sensitive to certain types of error in the spatial morphology of the Poissonian templates. Any attempt to claim robust evidence for PSs with other choices of the background templates and/or ROI will need to make the case that such systematic effects are unimportant.

\begin{figure}[t!]
\includegraphics[width=0.49\textwidth]{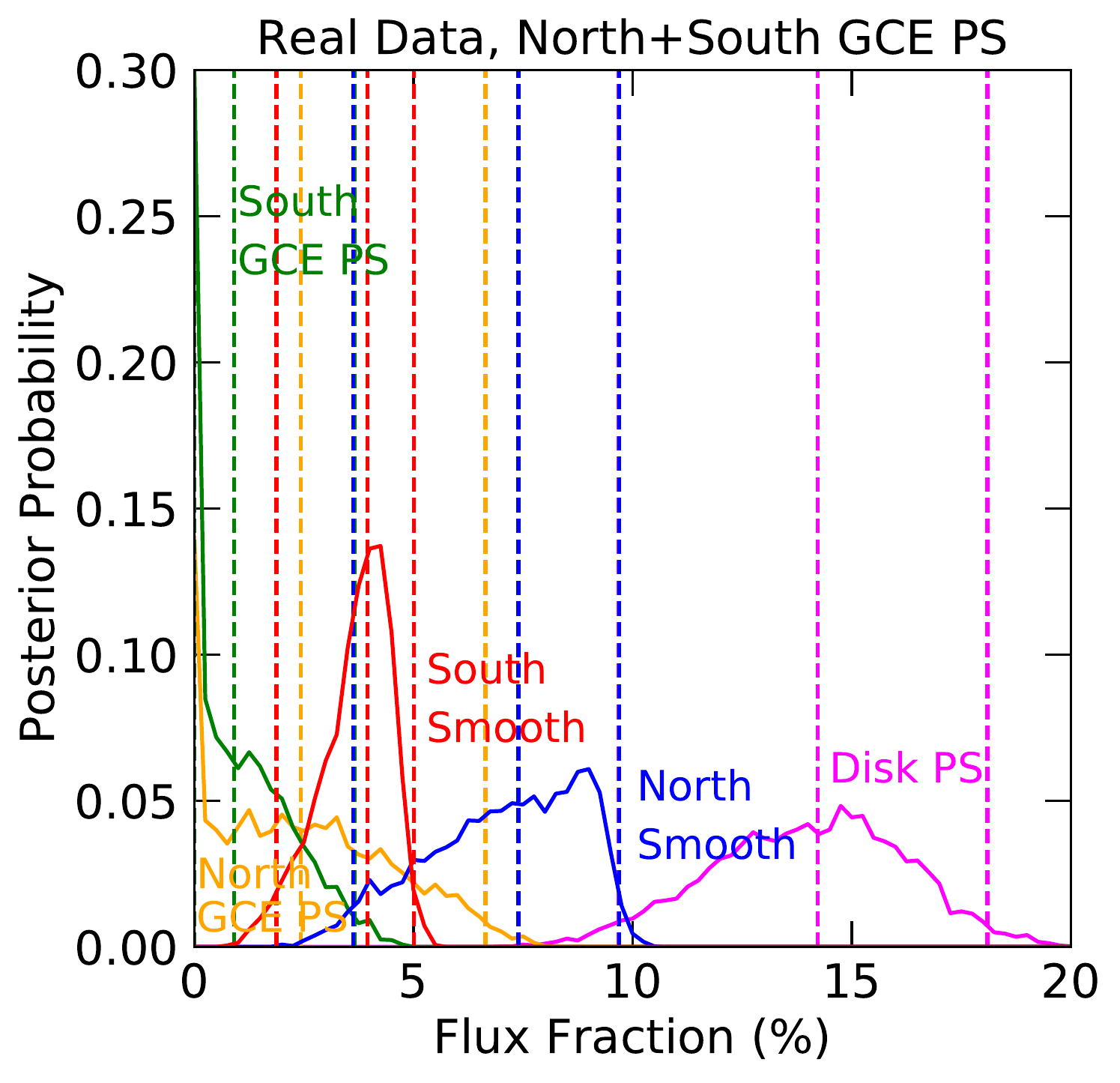}
\caption{Posterior distributions for the flux fractions (within the ROI) associated with various templates in the real data, when the north and south smooth and PS GCE templates are permitted to float separately. This fit is with \texttt{p6v11} diffuse model. With this additional freedom, the fluxes for the PS templates are consistent with zero.}
\label{fig:a_flux}
\end{figure}

\begin{table}[t]
\centering
\renewcommand{\arraystretch}{1.5}
\begin{tabular}{cccc}
\hline
\textbf{GCE Signal} & \hspace{5mm}\texttt{Model A} &\hspace{2mm}\ \texttt{Model F} & \hspace{2mm}\texttt{p6v11} \\ \hline
One template & \hspace{5mm}$4\times10^2$ &\hspace{2mm} $1$ & \hspace{2mm}$4\times 10^{15}$ \\ 
North+South  & \hspace{5mm}$6$ &\hspace{2mm} $1$ & \hspace{2mm}$7$  \\ \hline
\end{tabular}
\caption{Bayes factors in favor of GCE PSs. The ``GCE Signal'' column describes whether the GCE templates are assumed to be north-south symmetric (both smooth and PS), or subdivided into northern and southern halves and floated separately (both smooth and PS).}
\label{tab:bayesNFWreg}
\end{table}

We can check if introducing separate northern and southern templates artificially degrades the preference for PSs in simulations (due to extra unneeded degrees of freedom), when a GCE PS population is in fact present. We simulate a symmetric GCE PS population (with parameters taken from the fit to the real data, assuming symmetric GCE templates). We find that when analyzing these simulations with either symmetric GCE templates, or separate northern and southern templates, we obtain comparable Bayes factors
in favor of PSs when analyzing with symmetric or asymmetric PSs. That is, while we note that in some simulations the Bayes factors may be up to $1-2$ orders of magnitude lower when the PS templates are split in the asymmetric case, the drop in Bayes factors from using asymmetric PS templates are never approaching the factor of $10^{15}$ drop observed in the real data.
\\

\textbf{\textit{The Fermi Data Strongly Prefer Additional Template Freedom.}} 

We now determine whether a signal template with north-south asymmetry can provide a significantly better description of the GCE morphology. To address this question, we consider only the GCE Smooth template, which can be divided into different subregions; we do not include a GCE PS template. Our companion paper presents tests of other subdivisions beyond north-south asymmetry~\cite{PhysRevD.102.063019}.

Figure~\ref{fig:10degasym} shows the posterior fluxes for the 10$^\circ$ region, with only smooth GCE templates split into north-south components, for diffuse model \texttt{p6v11}. A clear asymmetry is found.

\begin{figure}[t]
\leavevmode
\centering
\includegraphics[width=0.49\textwidth]{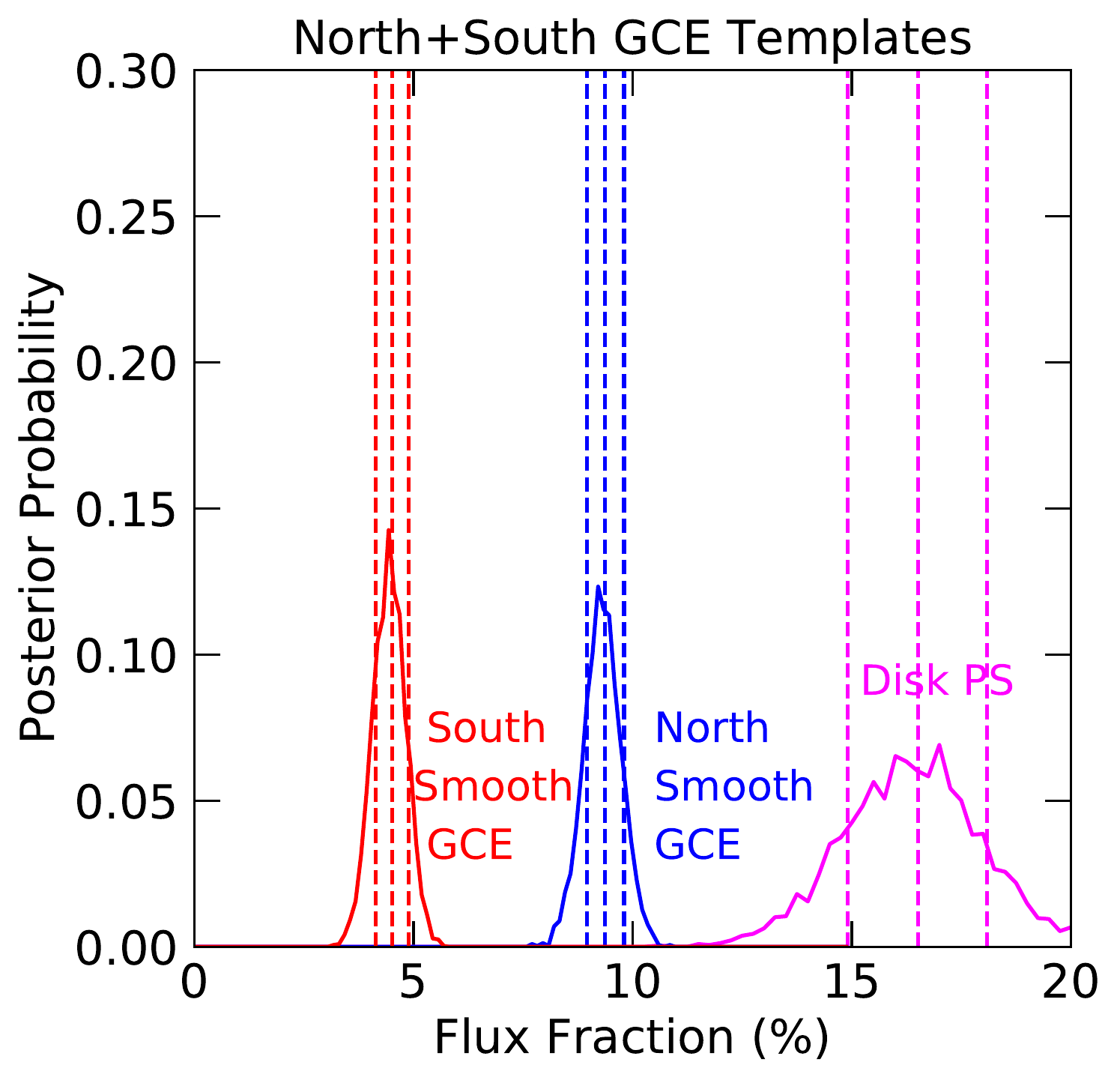}
\caption{Flux posteriors demonstrating the impact of providing a smooth GCE template with additional freedom, in the real data. This fit is with \texttt{p6v11} diffuse model. No GCE PSs are included in the fit; the smooth GCE template is divided into independent north and south regions. 
}
\label{fig:10degasym}
\end{figure}

Table~\ref{tab:bayesasym} summarizes our results for three different diffuse models. In all cases we find a significant preference for GCE asymmetry, with a particularly high Bayes factor of $\sim 10^{27}$ when using the \texttt{p6v11} diffuse model.

\begin{table}[b]
\centering
\renewcommand{\arraystretch}{1.5}
\begin{tabular}{cc}
\hline
\textbf{Diffuse Model}   &\hspace{5mm} \textbf{Bayes Factor for Asym GCE  }\\ \hline
\texttt{p6v11}  & $2\times10^{27}$   \\
\texttt{Model A} & $1\times10^{3}$ \\ 
\texttt{Model F}  &  $6\times 10^{4}$  \\ \hline
\end{tabular}
\caption{Preference for smooth GCE asymmetry with three diffuse models: \texttt{p6v11}, \texttt{Model A}, and \texttt{Model F}. Bayes factors are in favor of the scenario where the smooth GCE template is subdivided into independent north and south components, compared to the simpler scenario with a single smooth GCE template. }
\label{tab:bayesasym}
\end{table}

\begin{figure*}[t]
\leavevmode
\centering
\subfigure{\includegraphics[width=0.33\textwidth]{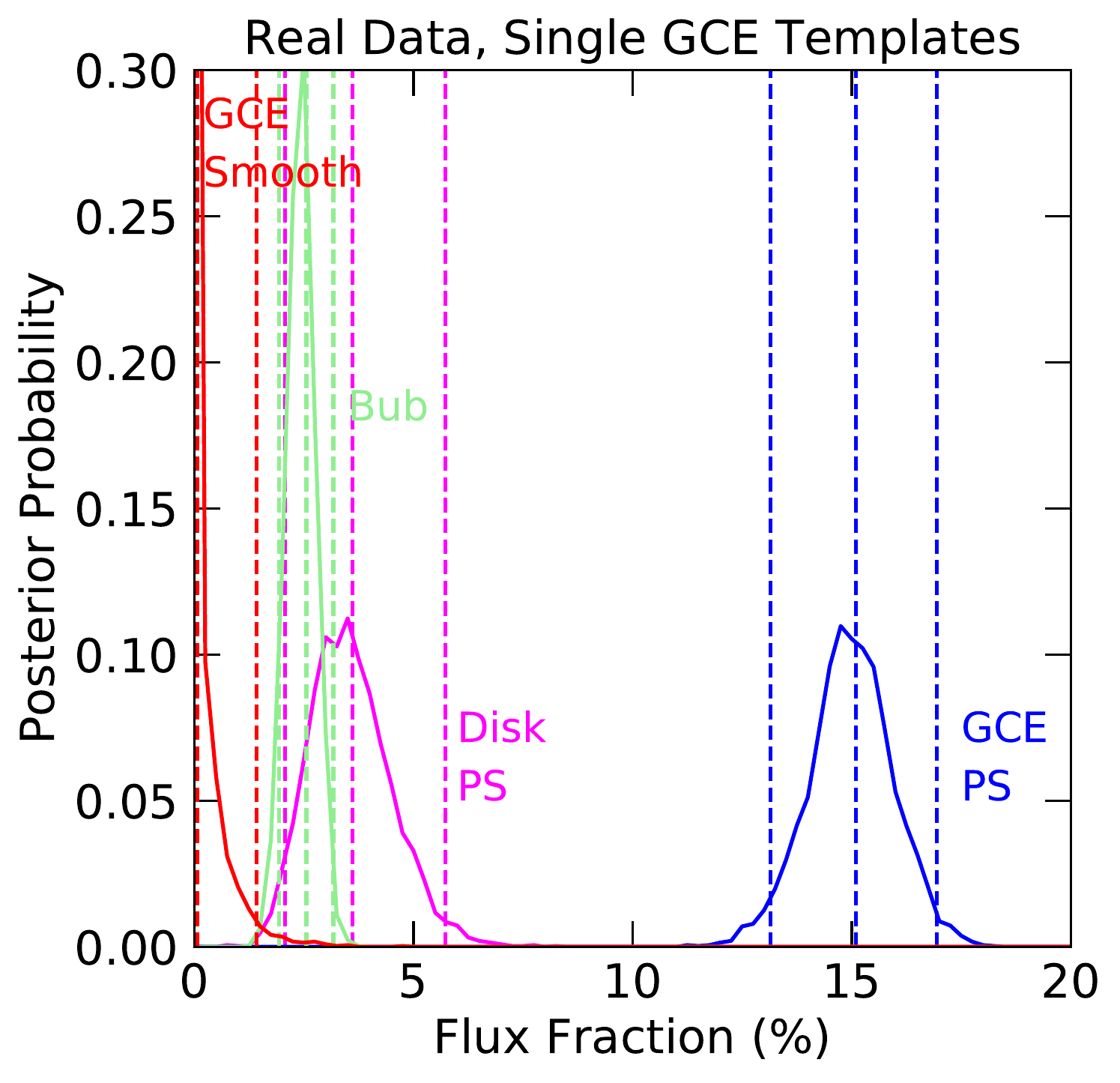}}
\subfigure{\includegraphics[width=0.32\textwidth]{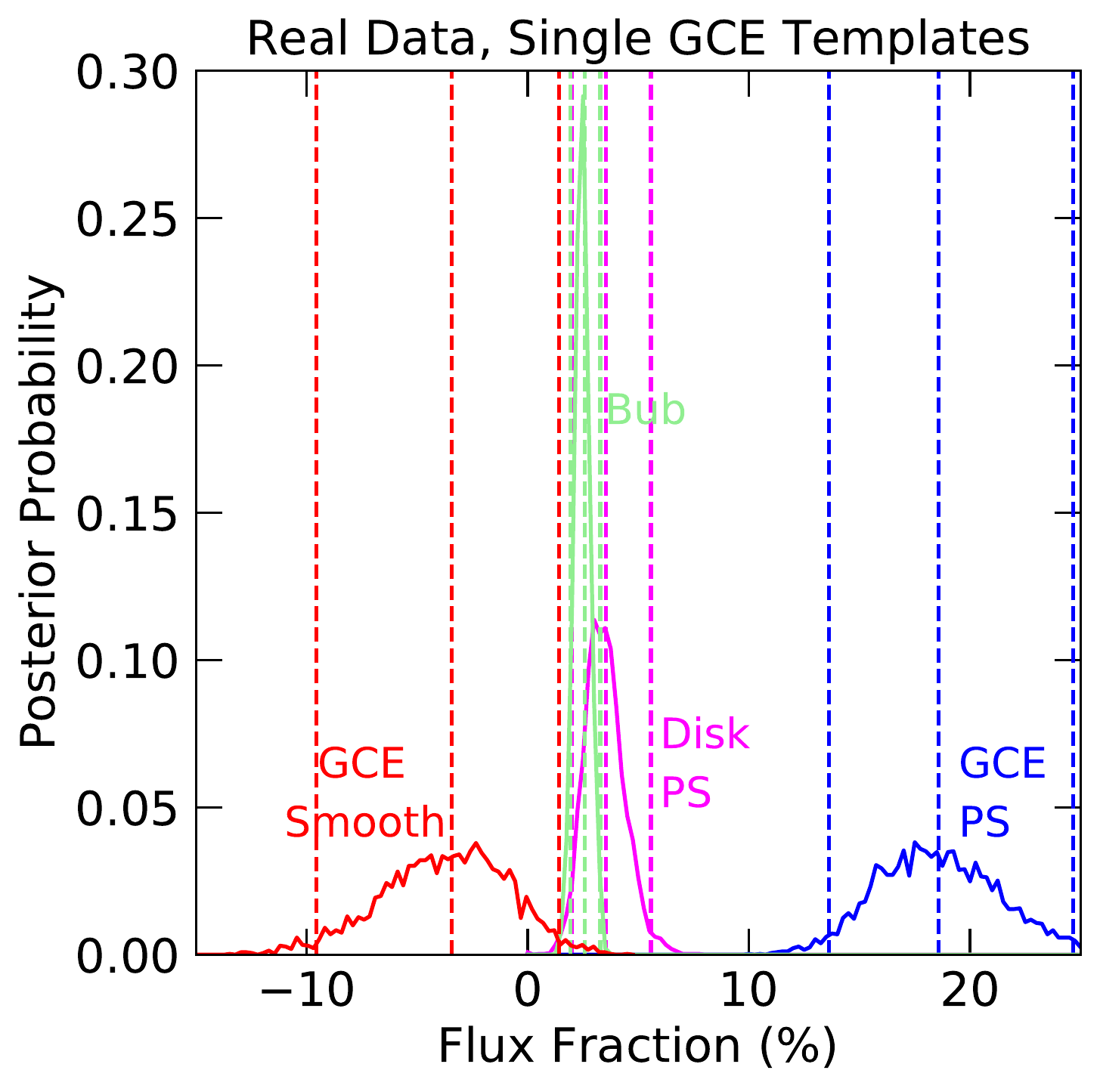}}
\subfigure{\includegraphics[width=0.32\textwidth]{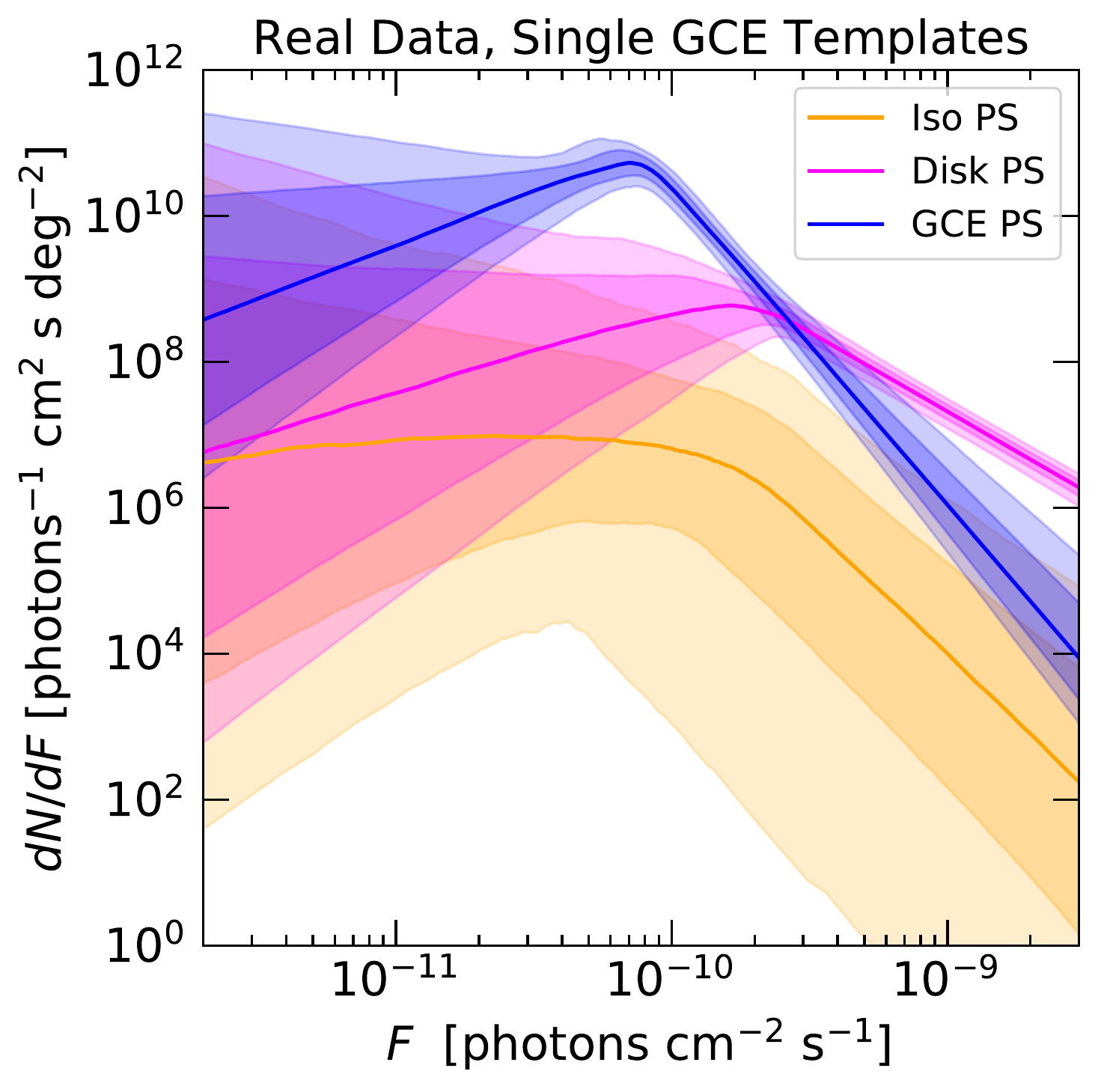}}\\
\subfigure{\includegraphics[width=0.33\textwidth]{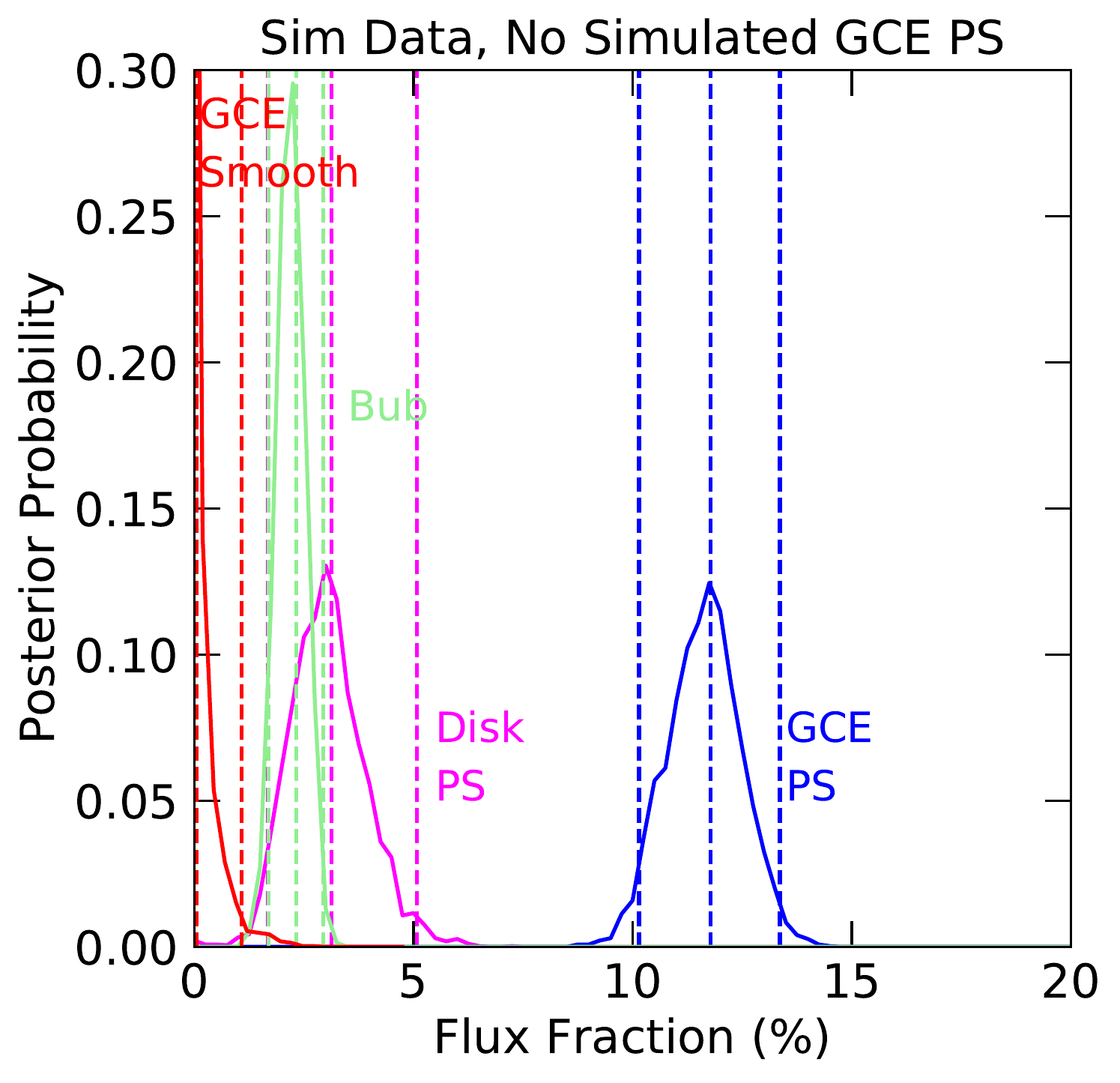}}
\subfigure{\includegraphics[width=0.32\textwidth]{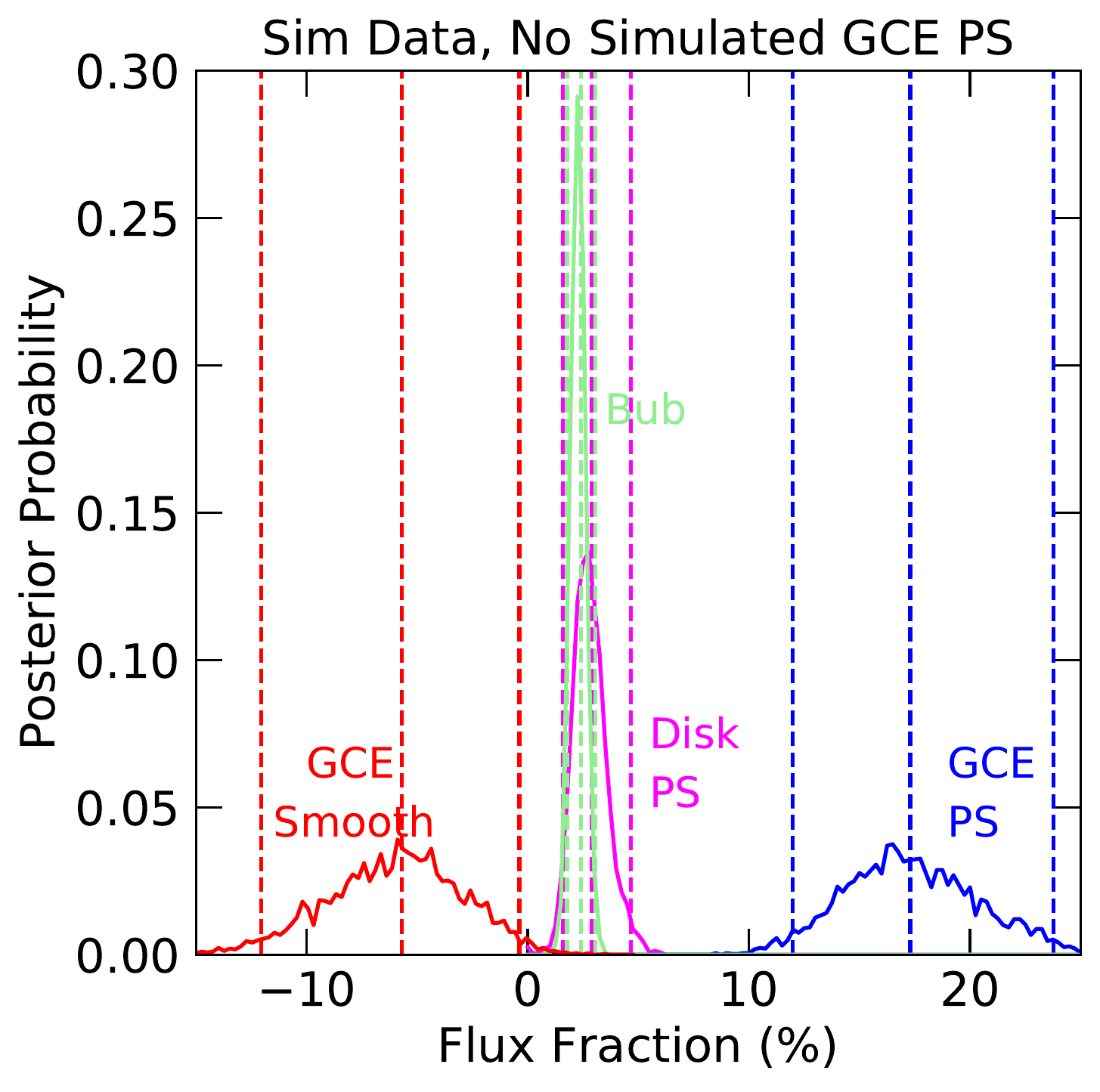}}
\subfigure{\includegraphics[width=0.32\textwidth]{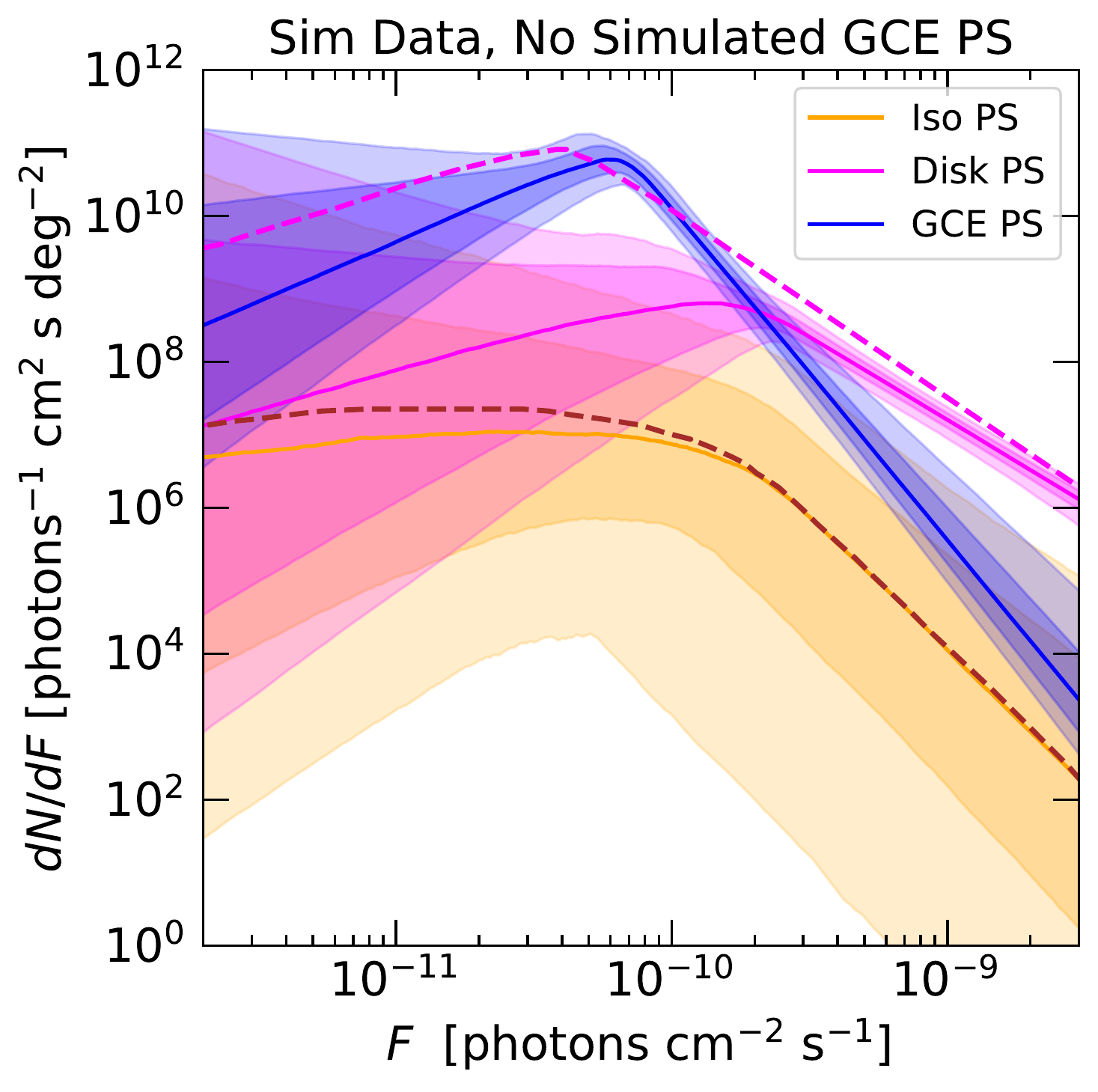}}\\
\caption{Comparison of real (\emph{top row}) and simulated (\emph{bottom row}) data; in all cases the analyses used symmetric GCE templates (smooth and PS). The simulated dataset contains a smooth asymmetric GCE and no GCE PSs.
\textbf{Left column:} Flux posteriors for various templates in the fit where the GCE Smooth component is constrained to have positive coefficient. \textbf{Middle column:} Flux posteriors for various templates in the fit where the GCE Smooth component is allowed to float to negative values. \textbf{Right column:} SCF corresponding to the left column. The dashed lines on the bottom SCF plot are the simulated SCF for Disk PS (pink) and Iso PS (brown).}
\label{fig:10deg}
\end{figure*}

The subdivided GCE Smooth template is preferred to the case with a single GCE Smooth + single GCE PS template with a Bayes factor of $\sim5\times10^{11}$ for \texttt{p6v11}. Thus subdividing the GCE Smooth template provides a much better description of the data than adding a symmetric GCE PS template to the model.

Other changes to the diffuse model (see Supplemental Materials), or changes to the other background templates (e.g. the \textit{Fermi} Bubbles) could potentially affect the inferred GCE asymmetry. We emphasize that we are not claiming that the physical mechanism giving rise to the GCE must possess a north-south asymmetry, just that within the set of background models tested, such an asymmetry provides a markedly better description of the data. Note it does not appear with all background models in larger ROIs; the generality of our results are discussed in a companion paper~\cite{PhysRevD.102.063019}.

With the observation that north-south asymmetry is preferred in the GCE, and that affording this freedom to the GCE removes the preference for PSs, we now investigate if we can explain this picture in simulated data.\\ \\

\noindent
\textbf{\textit{Evidence in Simulations: Unmodeled Asymmetry can lead to Spurious Point Source Detection.}}

We now simulate 100 realizations of the scenario where the GCE Smooth template has northern and southern components with differing normalization (no other templates are subdivided), using the posterior medians from the previous section (i.e. based on Fig.~\ref{fig:10degasym}). No GCE PS template is simulated. 
We then analyze these realizations with the pipeline used in previous works~\cite{Lee:2015fea,Leane:2019xiy,Chang:2019ars}, i.e. where the GCE model allows a symmetric PS template and a symmetric smooth template.

Figure~\ref{fig:10deg} (lower panels) shows our results for a realization chosen to resemble the real data (see the Supplemental Material for the full range of realizations).  We find that fitting the asymmetric GCE with purely symmetric templates drives the GCE Smooth template coefficient to zero (when the prior enforces that this coefficient must be non-negative), and produces a strong preference for a spurious population of GCE PSs, which approximately absorbs the flux of the simulated smooth GCE.  Ref.~\cite{Leane:2019xiy} previously demonstrated that systematic mismodeling can cause the GCE Smooth template coefficient to be driven to unphysical negative values, and that the same behavior can be observed in the real data. Thus, in the central panels of Fig.~\ref{fig:10deg} we relax the prior on the GCE Smooth coefficient to allow negative values, and observe that the unmodeled asymmetry again has the effect of driving the GCE Smooth coefficient negative. 

In the upper panels of Fig.~\ref{fig:10deg} we show the results of the identical analyses on real data in our ROI. The similarity between the real data and the selected realization is striking, both in the flux fractions and in the SCF for the inferred GCE PS population (which in the simulated data is certainly spurious). In particular, we see that the bright end of the SCF matches closely between the two analyses; that is, the NPTF can incorrectly infer the existence of even relatively bright PSs (and this behavior is generic across realizations). The Bayes factors in favor of GCE PSs, in the real data and selected realization, are $4\times10^{15}$ and $4\times10^{12}$ respectively; both are within the range spanned by our 100 realizations, but at the high end (see Supplemental Material for details). 
We find similar results for these tests with the alternate diffuse model, \texttt{Model A}; see our companion paper~\cite{PhysRevD.102.063019} for details.

In both real and simulated data, the amount of flux attributed to disk PSs drops precipitously when the GCE is forced to be north-south-symmetric. We observe in the simulated realization that disk PSs are being incorrectly reconstructed as GCE PSs. Therefore, failing to account for the asymmetry (or other mismodeling) can cause mis-allocation of real PSs to the wrong morphology, as well as inducing spurious PSs.\\

\noindent
\textbf{\textit{Conclusions and Outlook.}}

We have investigated the role of errors in the signal template in the preference for a large PS contribution to the GCE. We find that additional freedom in the morphology of the GCE signal template -- specifically, a north-south asymmetry -- is preferred by the data, and omitting it can lead to a spurious preference for PSs, and an artificial oversubtraction of the smooth GCE component.

We found in simulated data that if the signal has a north-south asymmetry, but is fitted with a north-south-symmetric template, then the fit reconstructs a spurious GCE PS population at high significance, in a large majority of realizations. Furthermore, the spurious PSs can be quite bright, well above the degeneracy limit where a population of faint PSs becomes formally indistinguishable from a smooth signal, and providing a steeply peaked SCF very similar to that found in NPTF analyses of the real data.

The prospect of mismodeling-induced systematics has always been a concern for NPTF analyses \cite{Lee:2015fea, Chang:2019ars, Leane:2019xiy}, but previous studies have indicated that a wide range of diffuse emission models can give broadly consistent results. Our results explicitly demonstrate, for the first time, that in the region where the GCE is brightest ($10^\circ$ radius ROI), with the diffuse models used in previous analyses, the NPTF-based evidence for PSs disappears given a more flexible signal model, and the data are better explained by north-south asymmetry of a smooth GCE.

We emphasize that we do not claim this asymmetry is a robust intrinsic property of the GCE, as it is plausible it arises from cross-talk with other mismodeled templates. However, if future analyses were to demonstrate that the central part of the GCE does indeed possess a robust and pronounced north-south asymmetry, that would strongly constrain possible GCE origins. We expect it would likely be challenging to obtain a large asymmetry in the context of DM annihilation.

Instead, we argue that with our current background templates and ROI, the fit prefers this asymmetry over the GCE point-source hypothesis, and not including it in the model then leads to a spurious PS detection. The reason is that a PS population can more easily accommodate a large pixel-to-pixel variance, as we discuss in detail in our companion paper~\cite{PhysRevD.102.063019}. This strikes a note of caution for future NPTF studies, as many causes of increased variance (i.e. mismodeling) could generate spurious PS signals. It will be important to consider ways to mitigate this issue, or demonstrate that it is irrelevant in a given analysis.

More generally, this result serves as a warning that even a highly-significant preference for PSs, with an inferred SCF that includes relatively bright PSs, need not be reliable. In our companion paper~\cite{PhysRevD.102.063019}, we develop a simplified analytic description of how signal mismodeling drives a preference for spurious PSs, explore the behavior of our results under variations on this analysis, and discuss the broader implications of these findings for past and future NPTF studies.\\

\textbf{\textit{Acknowledgments.}}
We thank M. Buckley, M. Buschmann, L. Chang, G. Collin, R. Crocker, D. Curtin, D. Finkbeiner, P. Fox, D. Hooper, S. Horiuchi, T. Linden, M. Lisanti, O. Macias, S. McDermott, S. Mishra-Sharma, S. Murgia, K. Perez, N. Rodd, B. Safdi, T. Tait, and J. Thaler for helpful discussions. We thank the \textit{Fermi} Collaboration for the use of \textit{Fermi} public data. The work of RKL was performed in part at the Aspen Center for Physics, which is supported by NSF grant PHY-1607611.  RKL and TRS are supported by the Office of High Energy Physics of the U.S. Department of Energy under Grant No. DE-SC00012567 and DE-SC0013999, as well as the NASA Fermi Guest Investigator Program under Grant No. 80NSSC19K1515.

\bibliography{annulus}

\clearpage
\newpage
\maketitle
\onecolumngrid
\begin{center}
\textbf{\large Spurious Point Source Signals in the Galactic Center Excess}

\vspace{0.05in}
{ \it \large Supplementary Material}\\ 
\vspace{0.05in}
{Rebecca K. Leane and Tracy R. Slatyer}
\end{center}
\onecolumngrid
\setcounter{equation}{0}
\setcounter{figure}{0}
\setcounter{section}{0}
\setcounter{table}{0}
\setcounter{page}{1}
\makeatletter
\renewcommand{\theequation}{S\arabic{equation}}
\renewcommand{\thefigure}{S\arabic{figure}}
\renewcommand{\thetable}{S\arabic{table}}
\vspace{-3mm}

\section{Parameters and priors}

Table~\ref{tab:sim_values_one} describes the parameters used in simulations throughout this work, taken from the posterior medians in the fit to the real data when the smooth GCE template is broken into independently-floated northern and southern components, and no template for GCE PSs is included. When creating simulated data, we first perform a fit on the real data and calculate the posterior medians for the various model parameters; we then simulate data based on those parameter values.

Table~\ref{tab:priors} details the priors used in our analyses. $A_\text{template}$ denotes the coefficient of the template in question. The SCF for non-Poissonian templates is parameterized as $dN/dF = A (F/F_b)^{-n_2}$ for $F < F_b$, $dN/dF = A (F/F_b)^{-n_1}$ for $F \ge F_b$. By convention we state our prior on $F_b$ in terms of the average counts at the break; the conversion factor from $F_b$ to the average counts $S_b$ is the average exposure in the ROI. In our default dataset, the average exposure is $2.79 \times10^{11}$ cm$^2$s.

\begin{table*}[h]
\centering
\renewcommand{\arraystretch}{1.5}
\setlength{\tabcolsep}{5.2pt}
\begin{tabular}{cc}
\hline
\multicolumn{2}{c}{\textsc{Simulation Parameters}}\Tstrut\Bstrut \\
\hline
Parameter & Simulation Value\Tstrut\Bstrut \\
\hline 
\hline
$\log_{10}A_\text{iso}$   & ${-1.49}$  \Tstrut\Bstrut \\ 
$\log_{10}A_\text{dif}$   & ${1.15}$  \Tstrut\Bstrut \\
$\log_{10}A_\text{bub}$  & ${-1.20}$ \Tstrut\Bstrut \\ 
$\log_{10}A_\text{GCE}^\text{north}$  & ${0.67}$ \Tstrut\Bstrut \\ 
$\log_{10}A_\text{GCE}^\text{south}$  & ${0.34}$ \Tstrut\Bstrut \\ 
$\log_{10}A_\text{PS}^\text{disk}$  & $-1.53$  \Tstrut\\
 $S_b^\text{disk}$    & $12.95$    \\
 $n_1^\text{disk}$      & $2.55$    \\
 $n_2^\text{disk}$      & $-1.18$  \\
 $\log_{10}A_\text{PS}^\text{iso}$ & $-4.63$  \Tstrut\\
 $S_{b,1}^\text{iso}$     & $31.06$    \\
 $n_1^\text{iso}$    & $3.60$    \\
 $n_2^\text{iso}$      & $-0.52$  \\
\hline
\hline
\end{tabular}
\caption{Parameter values used to generate the simulated data from Poissonian and non-Poissonian templates for the case where the smooth GCE template is broken into northern and southern components which are floated independently (normalizations controlled by $A_\text{GCE}^\text{north,south}$), and no GCE PS template is included in the simulation. These values are taken from the posterior medians in the corresponding fit to real data.}
\label{tab:sim_values_one}
\end{table*}

\begin{table*}[h!]
\renewcommand{\arraystretch}{1.5}
\setlength{\tabcolsep}{5.2pt}
\begin{center}
\begin{tabular}{ c  c c  }
\hline
\multicolumn{3}{c}{\textsc{Prior Ranges}}\Tstrut\Bstrut		\\  
\hline Parameter 	 & \textit{Fermi} \texttt{p6v11} diffuse model & Models A, F \Tstrut\Bstrut \\
\hline 
\hline
$\log_{10}A_\text{iso}$   & $[-3,1]$ & $[-3,1]$  \Tstrut\Bstrut \\ 
$\log_{10}A_\text{dif}$   & $[0,2]$  & $-$ \Tstrut\Bstrut \\
$\log_{10}A_\text{GCE}^\text{north}$   & $[-3,1]$ & $[-3,1]$  \Tstrut\Bstrut \\ 
$\log_{10}A_\text{GCE}^\text{south}$   & $[-3,1]$ & $[-3,1]$  \Tstrut\Bstrut \\ 
$\log_{10}A_\text{GCE}$ & $[-3, 1]$  & $[-3, 1]$  \Tstrut\Bstrut	\\
$\log_{10}A_\text{ics}$   & $-$  & $[-2,2]$ \Tstrut\Bstrut \\
$\log_{10}A_\text{pibrem}$   & $-$  & $[-2,2]$ \Tstrut\Bstrut \\
$\log_{10}A_\text{bub}$  & $[-3, 1]$ & $[-3, 1]$\Tstrut\Bstrut \\ 
$\log_{10}A_\text{PS}$  & $[-6, 1]$ & $[-6, 1]$ \Tstrut\\
$S_b^\text{PS}$      & $[0.05 ,80]$  & $[0.05 ,80]$  \Tstrut\Bstrut \\
$n_1^\text{PS}$     & $[2.05, 5]$  & $[2.05, 5]$  \Tstrut\Bstrut  \\
$n_2^\text{PS}$      & $[ -3 ,1.95]$  & $[ -3 ,1.95]$   \Bstrut\\
\hline
\hline
\end{tabular}
\end{center}
\caption{Parameters and associated prior ranges used in all analyses unless explicity stated otherwise in the text. If the GCE Smooth template is permitted to float negative, it is analyzed with a prior range of $A_{\rm GCE}=[-9,9$].}
\label{tab:priors}
\end{table*}

\section{Splitting the Diffuse Model into North+South Templates}

If any asymmetry -- by which we mean a statistically significant preference for differing coefficients for the GCE across different subregions -- is detected, it is possible that this asymmetry might reflect errors in one or more of the background templates rather than a true asymmetry in the GCE. In particular, as the model for Galactic diffuse emission dominates the total photon flux, it is plausible that a slight mismodeling in the degree of asymmetry in the diffuse emission could induce a false preference for asymmetry in the GCE. As a first test of this possibility, we also explore the effect of subdividing the Galactic diffuse emission template into northern and southern halves, and letting their coefficients float independently, in combination with either a single GCE template or separate north/south GCE templates. All other templates are floated with a single coefficient over the full ROI.

Table~\ref{tab:bayesdiff} is a modified version of Table~II in the main text, adding information about floating the diffuse template in north-south pieces alongside the GCE template. We see that changing the Galactic diffuse emission model causes marked variations in the Bayes factor favoring asymmetry, but a preference remains in all cases we have tested. When the Galactic diffuse emission model is also subdivided into hemispheres, there is still a consistent preference for asymmetry in the GCE; the northern and southern GCE fluxes also both remain quite stable under this additional freedom.

We also compare the case where the GCE (but not diffuse model) is subdivided to the case where the diffuse model (but not GCE) is subdivided, for the \texttt{p6v11} diffuse template. In this case we find a Bayes factor favoring the GCE asymmetry of $\sim10^{10}$.

Lastly, we also compare the case where only the \textit{Fermi} Bubbles are allowed to float separately in north-south regions. While we find this improves the fit compared to when no templates are afforded north-south freedom at all (with a Bayes factor of $\sim10^{11}$), it does not impact the fit more than the diffuse model or GCE template being allowed north-south freedom.

\begin{table}[t]
\centering
\renewcommand{\arraystretch}{1.5}
\begin{tabular}{|c|c|c|}
\hline
\textbf{\begin{tabular}[c]{@{}c@{}} Diffuse\vspace{-2mm} \\ Model \end{tabular}}  & \begin{tabular}[c]{@{}c@{}}\textbf{Bayes Factor for} \vspace{-2mm} \\ \textbf{Asym GCE}\end{tabular}& \begin{tabular}[c]{@{}c@{}}\textbf{Bayes Factor for} \vspace{-2mm} \\ \textbf{Asym GCE} \vspace{-2mm}  \\ \textbf{Alongside Asym Dif}\end{tabular} \\ \hline\hline
\texttt{p6v11}  & $2\times10^{27}$  & $5\times10^{8}$ \\ \hline \hline
\texttt{Model A} & $1\times10^{3}$ & $4\times 10^{2}$  \\ \hline
\texttt{Model F}  &  $6\times 10^{4}$ &  $50$  \\ \hline
\end{tabular}
\caption{As per Tab.~II in the main text, but including information about splitting the diffuse model. Preference for smooth GCE asymmetry is shown with three diffuse models: \texttt{p6v11}, \texttt{Model A}, and \texttt{Model F}. Bayes factors are in favor of the scenario where the smooth GCE template is subdivided into independent north and south components, compared to the simpler scenario with a single smooth GCE template. In the left column no other templates (diffuse, isotropic+PS, bubbles, disk PS) are subdivided; in the right column the diffuse model is also subdivided into independent north and south components.}
\label{tab:bayesdiff}
\end{table}

\section{Hypothesis Comparisons in Additional Analyses}

In this section, we briefly compare additional combinations of analyses involving GCE PSs, and a smooth asymmetric GCE. In Table I of the main text, we compared the scenario where the data were analyzed with smooth north-south-asymmetric GCE templates, with the scenario where additional north and south GCE PS templates were included in the analysis. 

Table~\ref{tab:bayesPSs} shows additional log-evidences compared to the main text, allowing for comparison of additional scenarios.  To obtain the Bayes factor of interest, the difference in the relevant log-evidences should be exponentiated. 

\begin{table}[t]
\centering
\renewcommand{\arraystretch}{1.5}
\begin{tabular}{cccc}
\hline
\textbf{GCE Templates} & \hspace{5mm}\texttt{Model A} &\hspace{2mm}\ \texttt{Model F} & \hspace{2mm}\texttt{p6v11} \\ \hline
Symmetric Smooth & \hspace{5mm}$-5001.3$ &\hspace{2mm} $-4992.1$ & \hspace{2mm}$-5051.6$ \\ 
Symmetric Smooth + Symmetric PSs & \hspace{5mm}$-4995.3$ &\hspace{2mm} $-4991.8$ & \hspace{2mm}$-5015.3$ \\ 
Asymmetric Smooth  & \hspace{5mm}$-4994.7$ &\hspace{2mm} $-4981.2$ & \hspace{2mm}$-4988.2$  \\ 
Asymmetric Smooth + Asymmetric PSs  & \hspace{5mm}$-4992.7$ &\hspace{2mm} $-4981.9$ & \hspace{2mm}$-4986.2$ \\
Asymmetric Smooth + Symmetric PSs  & \hspace{5mm}$-4990.0$ &\hspace{2mm} $-4979.4$ & \hspace{2mm}$-4980.4$  \\  \hline
\end{tabular}
\caption{Log-evidences for analyses on the real data, when the GCE is modeled with the combination of templates listed. Alongside these templates, the data are analyzed with disk and isotropic PSs, along with smooth Bubbles, diffuse, and isotropic templates. Note that the numbers quoted generally have uncertainties of about $\pm0.25$.}
\label{tab:bayesPSs}
\end{table}

We comment on the new cases of interest in this table: the scenario where a smooth asymmetric GCE is allowed in the fit, compared with the case where it is included alongside a \textit{symmetric} GCE PS template, for various diffuse models. We find that for the \texttt{p6v11} diffuse model, the Bayes factor in favor of PSs is still dramatically decreased compared to the case where symmetric PSs are compared to a scenario with only a symmetric smooth GCE. Symmetric templates yield a Bayes factor for PSs of about $10^{15}$, while in the presence of the asymmetric smooth template, the Bayes factor in favor of a symmetric PS template falls to about $10^3$. For Model F, the Bayes factors are still $\sim1$ (it is interesting to note that this model also provides the best overall fit in this ROI and energy range, albeit not by a large factor). For Model A, the evidence is also still decreased, but not as substantially: symmetric templates yield a Bayes factor in favor of PSs of about $400$, while in the presence of an asymmetric smooth template the Bayes factor in favor of symmetric PSs drops to about $100$. We note that while these comparisons still exhibit a drop in the Bayes factor favoring PSs (except for \texttt{Model F} which is already only $\sim1$) in the presence of a smooth asymmetric GCE component, the residual evidence for PSs is higher than when north-south-asymmetric PSs are included in the fit  (as per Table I of the main text). However, the results of Table I and Table~\ref{tab:bayesPSs} are approximately consistent with each other, given the behavior we observed in simulations. That is, as discussed in the main text, we saw in simulations that when a symmetric GCE PS population was analyzed with one north and one south PS template, the Bayes factor could drop by about $1-2$ orders of magnitude, presumably due to the extra degrees of freedom in the PS model due to the additional independent PS template.

Comparing the log-evidences of all the scenarios in Table~\ref{tab:bayesPSs}, we note that the asymmetric smooth + symmetric PS scenario appears to be the most-preferred scenario overall. We do not interpret this as robust evidence for GCE PSs, as we have already demonstrated that unmodeled smooth asymmetries and symmetric PSs can be readily confused by the fit, and it is quite plausible that including the symmetric PS template absorbs some residual unmodeled asymmetry (it would be surprising if our discontinuous north-south-asymmetric template precisely matched the true GCE). Furthermore, the overall evidence for the scenario with model F and only an asymmetric smooth GCE is not significantly worse than any of the scenarios containing PSs. However, it is possible that the residual PS preference could also be a hint of an additional, albeit faint, GCE PS signal under the smooth asymmetry. 

\section{Spread of Simulated Data Analysis Results} 

In this section we show the distribution of results for fits to 100 simulated realizations, for the scenario described in Fig.~3 of the main text, where the GCE is simulated as a smooth signal with a north-south asymmetry, and no GCE PSs are simulated. The fit includes symmetric smooth and symmetric PS templates for the GCE, as well as the standard set of background templates.

\begin{figure}[h!]
\leavevmode
\centering
\includegraphics[width=0.45\textwidth]{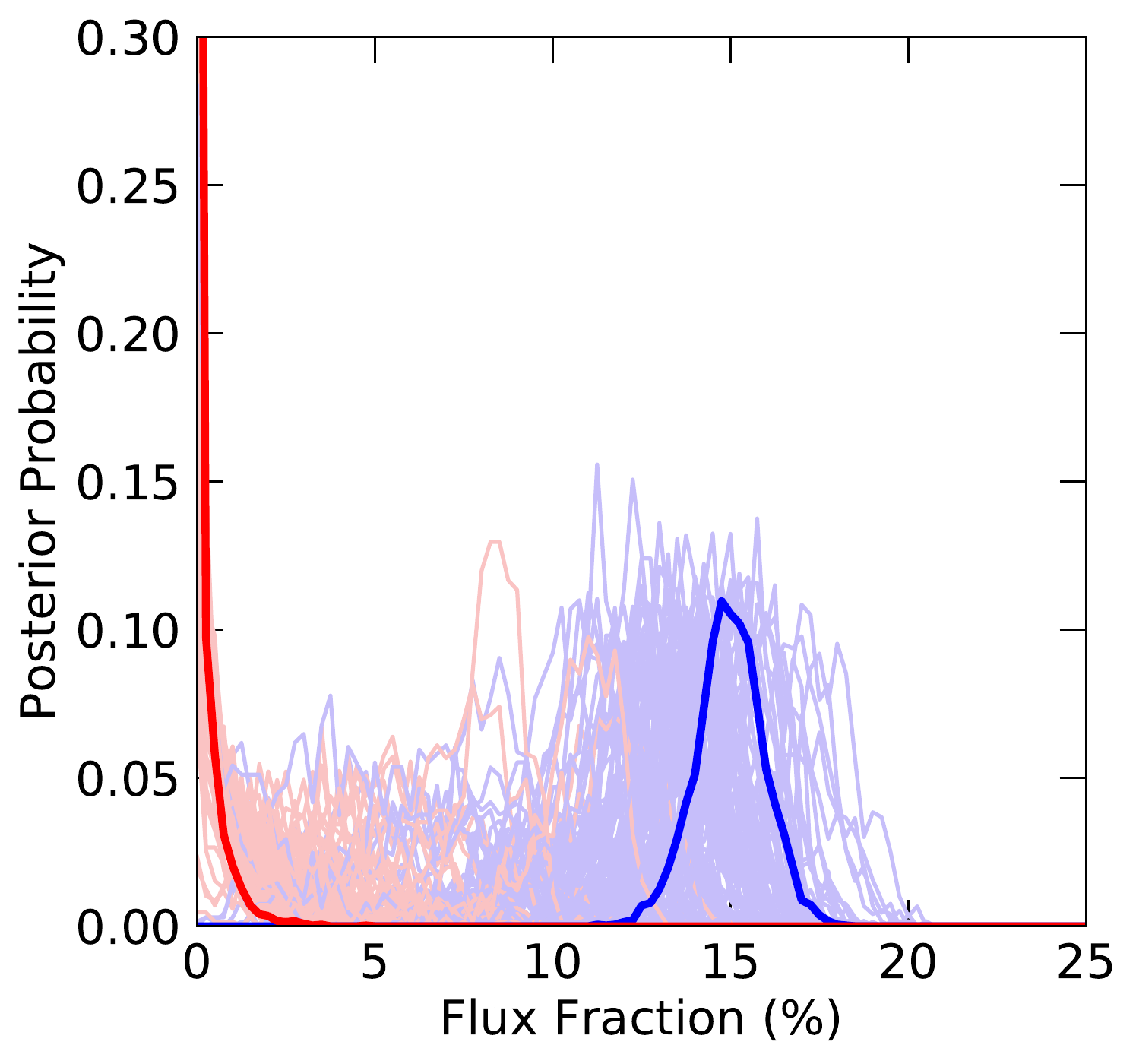} \hspace{3mm}
\includegraphics[width=0.44\textwidth]{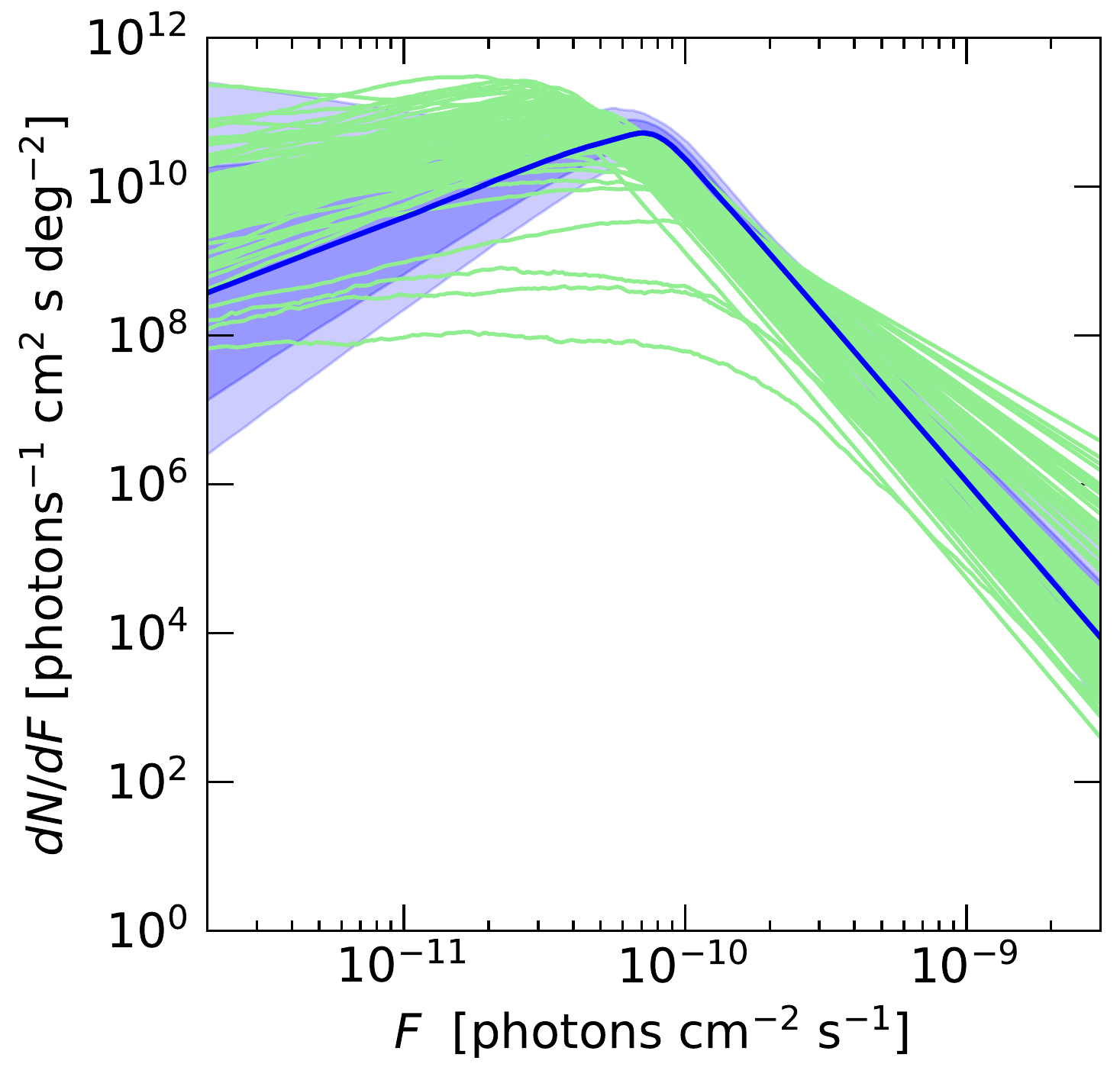}
\caption{Spread of analysis results on 100 simulated-data realizations, extending Fig.~3 of the main text. In all cases, the analyses use single GCE templates (smooth and PS) over the whole ROI. The simulated dataset is based on the best fit model (fluxes shown in Fig.~2) using separate Poissonian templates for the northern and southern GCE; no GCE PSs were simulated. \textbf{Left:} Flux fraction posteriors.  Fainter blue (red) lines correspond to the GCE PS (GCE Smooth) posteriors for simulated realizations, bold darker lines are the real data. \textbf{Right:} the SCF obtained in the real data using one symmetric GCE PS template is shown in blue, the posterior median values of the reconstructed SCFs for GCE PSs, in the simulations, are shown in green.}
\label{fig:10degspread}
\end{figure}

Figure~\ref{fig:10degspread} shows the spread of the posterior probability distributions for the GCE PS and GCE smooth flux fractions, and the SCF for the GCE PS population, compared to the posterior for the flux fraction and SCF in the real data (in fits where the GCE Smooth component is forced to be non-negative). We see that the results in real data are well within the band spanned by the simulations, and that it is very common for the fit to attribute similarly large flux fractions and similar SCFs to the (non-existent in the simulations) GCE PS population. 

Over these 100 realizations, the Bayes factor preference for PSs ranges from $\sim1-10^{20}$; note this means that signal mismodeling of this type \textit{can} generate very large Bayes factors in favor of PSs but also may not, depending on chance (thus otherwise-identical studies using different subsets of a dataset may obtain different results). The median Bayes factor (denoted BF) among the simulations is $\sim 2\times 10^5$, but the distribution in $\ln\text{BF}$ is broad, with an especially long tail at the high-Bayes-factor end. The realization shown in Fig.~3 of the main text has a Bayes factor of $4\times 10^{12}$, which is the 6th highest among the realizations; three realizations have Bayes factors higher than $10^{15}$. Thus, it is quite plausible that the Bayes factor for PSs found in real data is drawn from this distribution, but at the same time, it is possible that other deviations between the background models and the truth could be inflating the observed Bayes factor in real data.

Over the 100 realizations, GCE Smooth fluxes that are comparably negative to the results in the real data occur $\sim10\%$ of the time. The real data are thus consistent with being drawn from the simulation results, but again, it is possible that other sources of mismodeling are helping drive the GCE Smooth component slightly more negative, as we know that in larger ROIs, the GCE Smooth component in real data prefers a very negative value likely driven by mismodeling.

We also compare in simulations the scenarios where the GCE is $100\%$ smooth and symmetric (based on analysis of real data with no GCE PSs), and where the GCE is $100\%$ symmetric PSs (based on analysis of real data including both GCE Smooth and GCE PS templates), versus the scenario of Fig.~3 of the main text where the GCE is smooth and asymmetric (based on analysis of real data, with only a GCE Smooth template, where north and south are allowed to float independently). 

\begin{figure}[t]
\leavevmode
\centering
\includegraphics[width=0.45\textwidth]{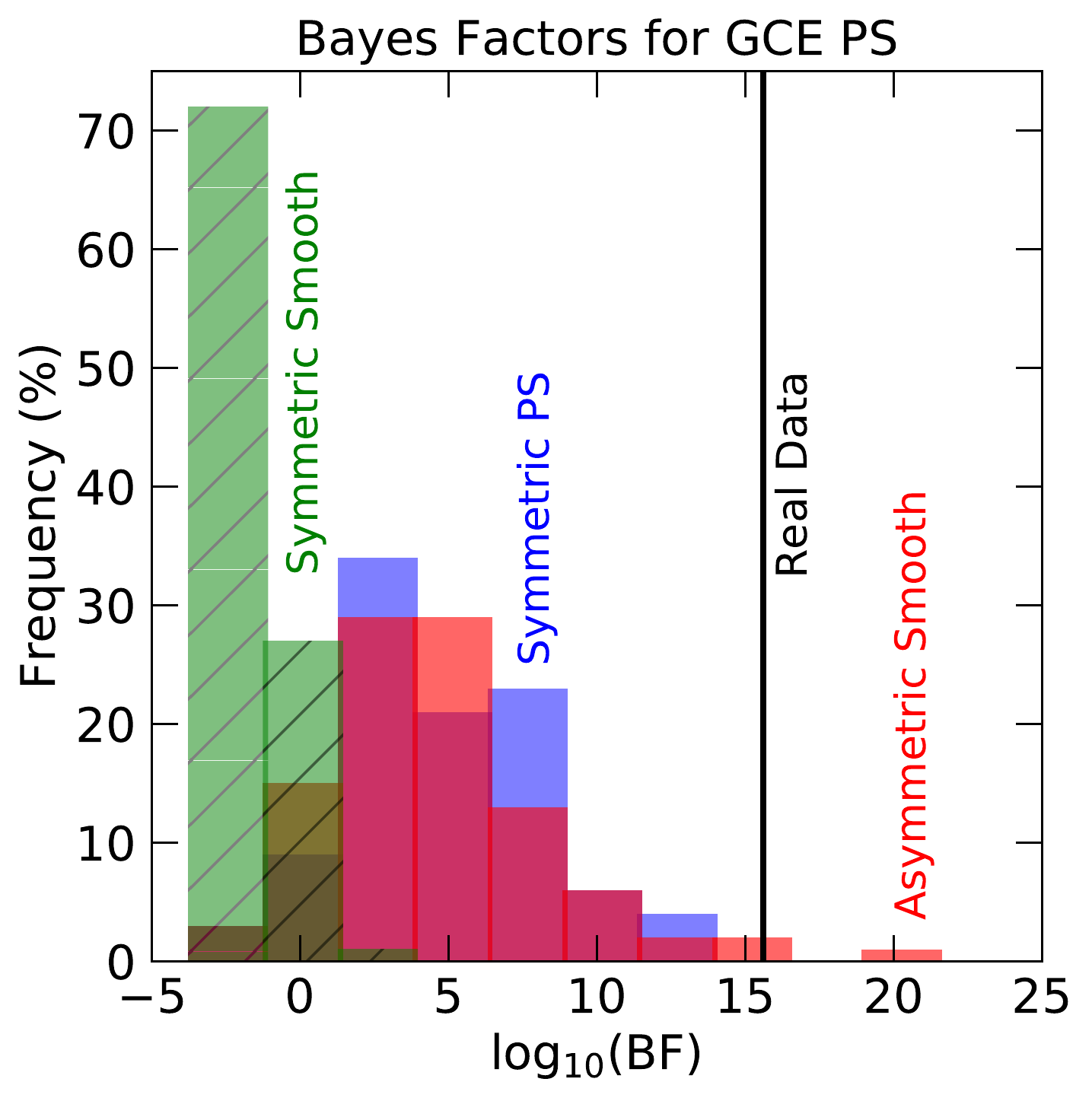}
\caption{Histogram of $\log_{10}$(BF) over 100 realizations for each of three simulated scenarios: (1) where the GCE is $100\%$ smooth and symmetric (parameters based on analysis of real data with no GCE PSs) (2) where the GCE is $100\%$ PSs (parameters based on analysis of real data including both GCE Smooth and GCE PS templates), and (3) where the GCE is smooth and asymmetric (parameters based on analysis of real data, with only a GCE Smooth template, subdivided into independent north and south components). The black line shows the $\log_{10}$(BF) value found for the real data with the same analysis. Note ``BF'' here is an abbreviation for Bayes factor.}
\label{fig:BF}
\end{figure}

Figure~\ref{fig:BF} shows the distribution of log$_{10}$(BF) for GCE PSs in each of these scenarios, where 100 simulations were performed for each dataset. We see that while the Bayes factor in the real data is on the high end compared to all the distributions of Bayes factor in simulated realizations, it is the most consistent with the simulations containing an asymmetric smooth GCE, followed closely by the simulations containing a symmetric PS-dominated GCE. We note that simulations with only a symmetric smooth GCE cannot produce Bayes factors comparable to what is observed in the real data, consistent with earlier studies \cite{Lee:2015fea,Chang:2019ars}; the mismodeling associated with the north-south asymmetry (or a similar effect) is needed to generate such a high-significance population of spurious PSs. However, the similarity of the Bayes factor distributions in the cases with an asymmetric smooth GCE or a symmetric GCE PS population means that Bayes factor is probably not a useful metric for discriminating between these scenarios.

Finally, note that in these analyses, the runs were only completed up to \texttt{nlive=100}, to speed up the large number of simulations. Running to higher \texttt{nlive} can often remove the tails of the posterior fluxes shown in Fig.~\ref{fig:10degspread}.

\section{Comparison with Source Catalogs}

\begin{figure}[t]
\leavevmode
\centering
\includegraphics[width=0.45\textwidth]{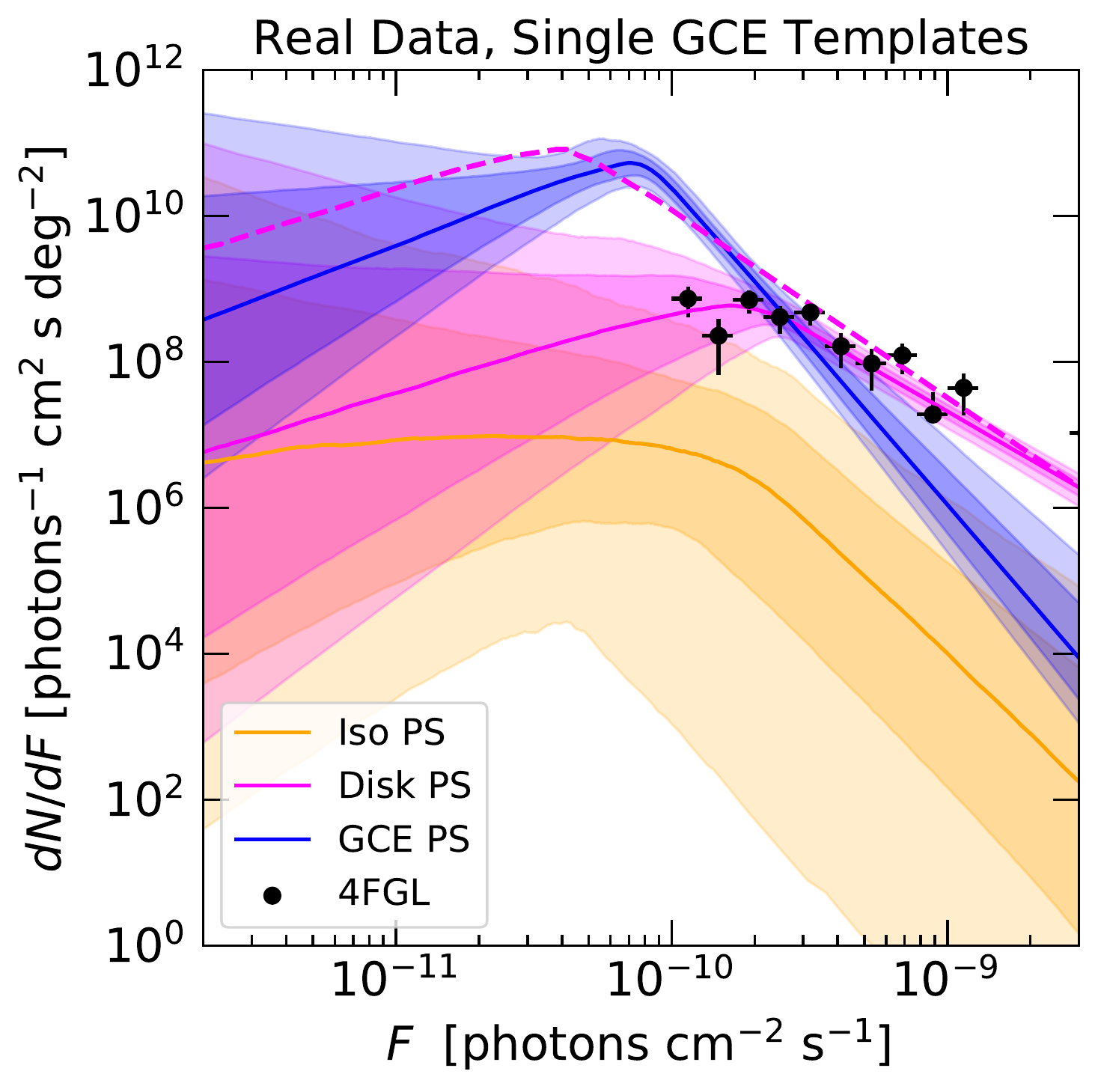}
\includegraphics[width=0.45\textwidth]{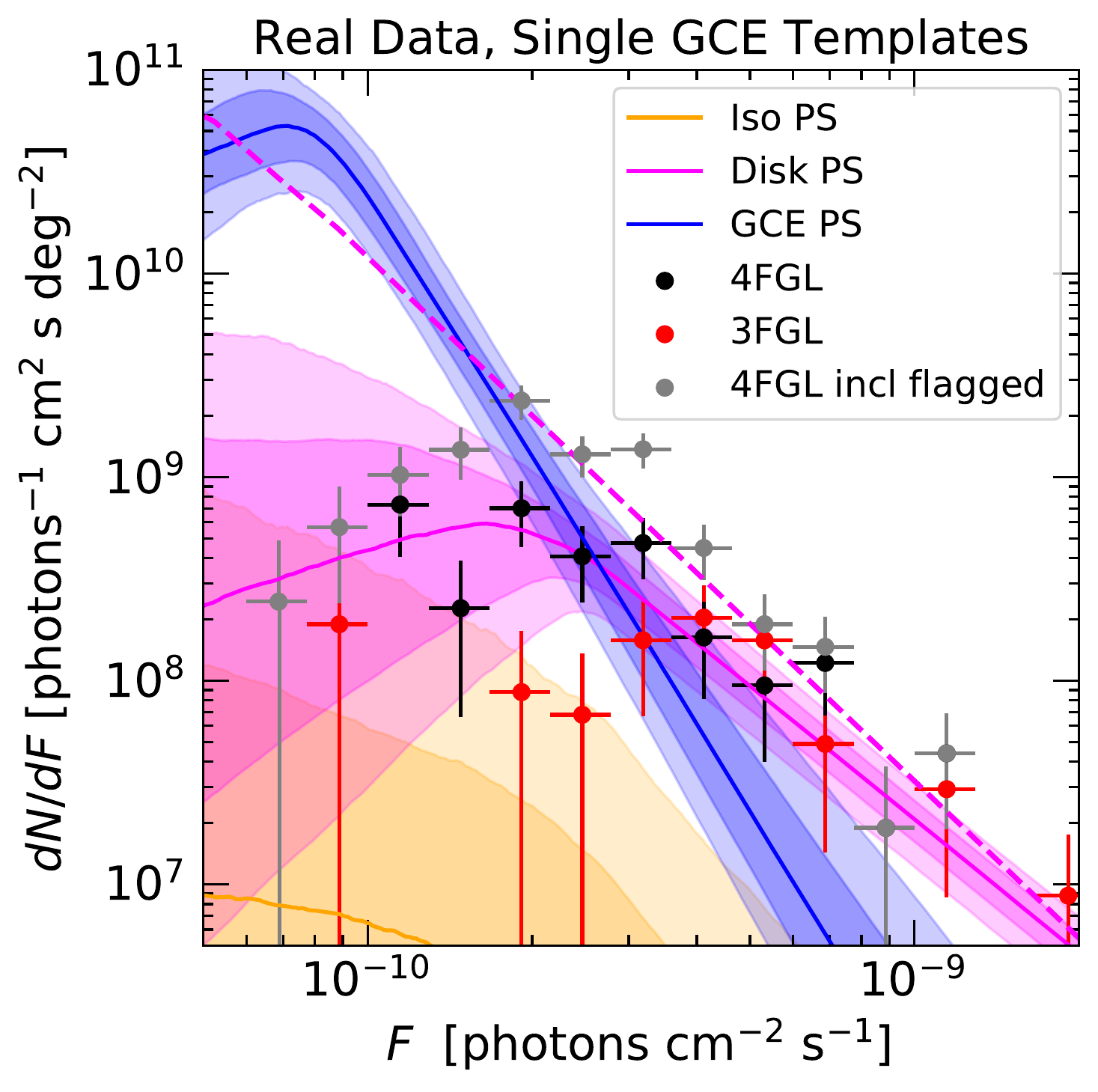}
\caption{SCF in the 10 degree region, when analyzing the real data with symmetric PS and Smooth templates.  \textbf{Left:} Overlay with the flux distribution of the 4FGL unflagged sources only  (\textit{black dots}). \textbf{Right:} Zoom in with overlay of the flux distribution for unflagged 3FGL sources (\textit{red dots}), unflagged 4FGL sources  (\textit{black dots}), and all 4FGL sources  (\textit{gray dots}). Horizontal bars on the data points denote the bin widths; the error bars on the $y$-axis correspond to $\sqrt{N}$ uncertainties on $N$, the number of sources per bin. The pink dashed line is the median posterior value of the disk SCF when performing the fit with a smooth asymmetric GCE plus backgrounds.}
\label{fig:4fgl}
\end{figure}

Figure~\ref{fig:4fgl} shows the SCF obtained in the 10 degree region, when analyzing the real data with symmetric GCE PS and GCE Smooth templates, overlaid with the fluxes for sources and source candidates from \textit{Fermi} point source catalogs, within the $10^\circ$ radius ROI. We also overlay the disk PS SCF obtained when the GCE is taken to be smooth and asymmetric (and hence there is no strong preference for GCE PSs). Some sources in each catalog are flagged as being of potential concern (due to e.g. sensitivity to the background modeling); we show the distribution of sources by flux for unflagged sources from the 3FGL catalog \cite{Acero:2015hja}, unflagged sources from the more recent 4FGL catalog \cite{Fermi-LAT:2019yla}, and all sources (flagged and unflagged) from the 4FGL catalog. This third sample extends down to the lowest fluxes, but may also contain a non-negligible number of spurious sources. Source fluxes for our energy band are obtained from the best-fit source spectrum models provided with the catalogs.

We observe that (as expected) the bright sources in all three samples are attributed to the disk PSs whether the GCE is assumed to be asymmetric or not. The faintest sources (in this energy band) in the 4FGL catalog have flux comparable to the crossover point where the GCE PSs would begin to dominate the PS population if they were present. We do not observe any obvious feature in the observed flux distribution that would correspond to a new GCE population, although the incompleteness of the catalogs at these low flux levels could potentially hide such a feature. The flux distribution for the fainter sources looks quite consistent with the disk SCF in the case where the GCE is assumed to be smooth and asymmetric.

\end{document}